\documentclass{vospwks2007}
\include{graphicx}
\title{Theoretical models in the Virtual Observatory}
\author[1,3]{C. Rodrigo}
\author[1,3]{E. Solano}
\author[1,3]{R. Guti\'errez}
\author[2,3]{M. Cervi\~{n}o}
\affil[1]{Laboratorio de Astrof\'{\i}sica Espacial y F\'{\i}sica Fundamental (LAEFF-INTA), P.O. 50727, E-28080 Madrid, Spain}
\affil[2]{Instituto de Astrof\'{\i}sica de Andaluc\'{\i}a (IAA-CSIC), Spain)}
\affil[3]{Spanish Virtual Observatory Thematic Network}

\begin{document}


\maketitle

\begin{abstract}
Although full interoperativity between theoretical and observational data
in the framework of the Virtual Observatory would be a very desirable achievement,
the current status of VO offers few approaches to handle theoretical models. 

TSAP (Theoretical Spectra Access Protocol) has been proposed as a tool to fill this void, providing a simple scheme to easily operate with this kind of data.

TSAP is useful not only for synthetic spectra but also for 
other types of theoretical data. As an example we show an Isochrone 
and Evolutionary Tracks server using TSAP. 

Finally, we pay special attention to the correct treatment of the credits an important issue in the field of theoretical models.

\end{abstract}

\section{Introduction}

Theoretical models are widely used in Astronomy. Synthetic spectra, for instance, can be used to infer the physical properties of an object by comparing its observed spectrum to a theoretical collection of spectra.
 
A number of libraries of theoretical models are presently available in the Internet. They  can be downloaded as a collection of data files with, in some cases, the help of a web form allowing a previous selection of the files of interest. The results are usually presented in different formats as ASCII or FITS files.

This scenario forces the user to perform a previous work in order to be able to compare theoretical and observational data. The situation is even worse if different sets of theoretical models, developed by different groups, are used. This lack of homogeneity makes it difficult to design automatic tools to simultaneously work with different models xand almost impossible to develop applications able to use the models on the fly.

One of the aims of the Virtual Observatory is to guarantee a full interoperativity not only between observational data but also between them and theoretical data. However, there are a number of issues that make it difficult to achieve this goal. One of them is the clear VO bias towards observational data. This can be seen, for instance, in the main protocols and standards already developed: SSAP, SIAP, ConeSearch,..., all of them require the object position and name as mandatory, parameters that are meaningless when dealing with theoretical models.

Actually, depending on the physical problem to be tackled, the particular approach taken by the author of the model or even his/her own personal preferences, the set of parameters used to characterize the model usually differs from one to another. This makes it difficult to define a general data model for theoretical data, a major topic for the IVOA and Euro-VO Theory Groups.

TSAP is being developed as a collaboration between ESAVO\footnote{http://esavo.esa.int} and SVO\footnote{http://svo.laeff.inta.es} as a way to make theoretical data easily available through the VO. It makes use of the already accepted standards although with some necessary modifications. 

\section{TSAP}

TSAP, originally based in the SSAP protocol, can be described as a dialogue between the client application and the model server based in three main steps: 

\begin{itemize}
\item What parameters define this model? what do they mean? and what values are allowed for each of them?

\item What files are available for a given range of those parameters?

\item How to obtain the files?

\end{itemize}

It is in the first step where the main differences with SSAP become remarkable. SSAP requires the object position (RA and DEC) together with a search radius as mandatory parameters, something that has no meaning when working with theoretical data. With TSAP the client first ask the server about the parameters than can be used to make the query. This is made in terms of an http query with the {\it format=metadata} parameter. On return, the server provides a VOTable that follows the SSAP specification and contains all the queriable parameters, their descriptions (human readable descriptions if possible) and, if desired, the allowed values or ranges of values. The client must be able to read this  metadata VOTable and build a form to make the real query. 

Here is an example of metadata VOTable:

\begin{scriptsize}

\begin{verbatim}
<votable>
...
<description>Stellar atmosphere
   model by...., version 1
   </description>
<param name=''teff_max'' 
    ucd=''phys.temperature.effective'' 
    units=''K'' 
    datatype=''float''>
  <description>Effective temperature 
        for the model in K</description>
  <values type=''actual'>
    <option value=1000/>
    <option value=1100/>
    <option value=1150/>
    <option value=3500/>
    <option value=4700/>
    <option value=5000/>
  </values>
</param>
...
</votable>
\end{verbatim}

\end{scriptsize}

This VOTable should be interpreted by the client application to build a form similar to that given in Figure 1.

\begin{small}
\begin{figure}[h]
\centering
\includegraphics[width=1.2\linewidth]{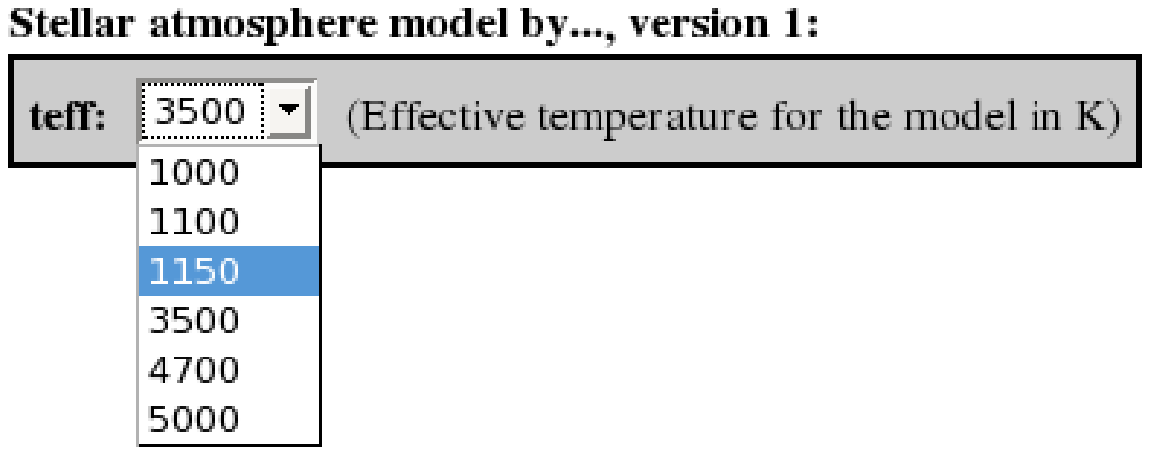}
\caption{A simple form generated during the TSAP client-server dialogue (see text for details).\label{fig:form4}}\end{figure}
\end{small}

Using this approach some data providers \footnote{http://laeff.inta.es/svo/theory/db2vo/html/}\footnote{http://ov.inaoep.mx/pgos3}\footnote{http://vo.obspm.fr/cgi-bin/siap/pegasehr.pl} are already offering theoretical spectra that can be understood by by VO-tools like VOSpec. 

A more detailed description of TSAP can be found in the TSAP page at the SVO web site.

\section{Accessing other theoretical data using TSAP}

Although TSAP was initially conceived for theoretical spectra, there is no reason why it cannot be used for any other kind of theoretical data. The possibility of asking the service about {\it what are the parameters} over which queries can be performed (the basis of TSAP) opens the VO to consult databases without the restriction of the name and/or a position in the sky.

We, at LAEFF, have developed a TSAP server\footnote{http://laeff.inta.es/svo/theory/draw/getiso.php?inises=guest} to give access to isochrones and evolutionary tracks. Complementary to this, we have developed a web application that uses TSAP to access and download the data. The application also implements visualization tools that allow the user to overplot different set of theoretical models to his/her own observational data (Figure 2).

At present, TSAP specifications only allow making a single query that can be mapped to a simple form. This limitation restricts the use of TSAP to services where the queriable parameter space is described as a n-dimensional cube. This situation can be easily alleviated by a simple extension of TSAP that allows recursive metadata queries defined by the service. See Cervino et al. (these proceedings) for more on this subject.

\begin{small}
\begin{figure}[h]
\centering
\includegraphics[scale=0.5]{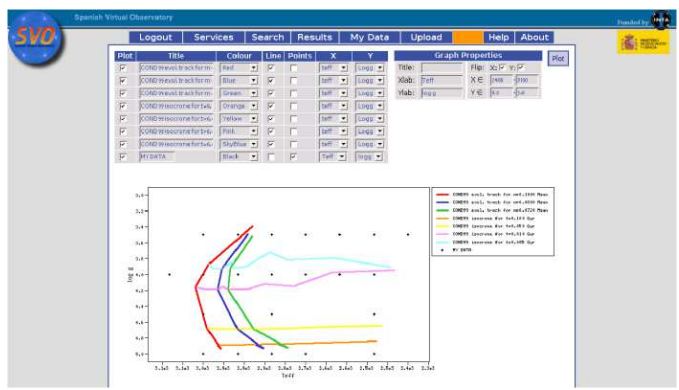}
\caption{Theoretical Isochrones obtained from the SVO Theoretical Data Server using TSAP are overplotted over an observational dataset.\label{fig:single3}}
\end{figure}
\end{small}

\section{Credits in the VO}
In the framework of the VO, data provenance (both theoretical and observational) is a fundamental issue. This is important not only for obvious scientific reasons but also to be able to adequately reference or acknowledge the scientific work and technical tools used.

VO services should provide this kind of information in their VOTables and VO applications should be able to keep track on them. In addition to include the information in the resulting VOTables, VO applications and services should articulate the mechanisms to display this information in such a way that allows its straightforward inclusion in a scientific publication (e.g. LateX file). This would benefit both the user and the theoretical data providers who would see it as an incentive to make their work available in the VO. This is the approach we are attempting to follow with the datasets contained in the SVO theoretical data server at LAEFF. 

\section*{Acknowledgements}

This research has made use of the Spanish Virtual Observatory supported from 
the Spanish MEC through grants AyA2005-04286, AyA2005-24102-E.

\end{document}